\documentclass[12pt,a4paper,notitlepage]{article}
\usepackage{amsmath}
\usepackage{graphicx}
\usepackage{amsfonts}
\usepackage{amssymb}
\begin{document}

\title{Self-consistent solution of the Kohn-Sham equations for systems with
inhomogeneous electron gas}
\author{A.Ya. Shul'man and D.V. Posvyanskii\\Institute of Radio Engineering and Electronics \\of the Russian Academy of Sciences, Moscow. ash@cplire.ru}
\maketitle

\section{INTRODUCTION}

The gas of the interacted electrons is usually described within Kohn-Sham
approximation by the set of Poisson and Schr\"{o}dinger equations with an
effective potential for the single-particle wave functions. The solution can
be obtained using many-step iteration procedure. The well known difficulty in
this task is that the wave functions obtained after every iteration step give
the distribution of electron density which is not correspond to the boundary
conditions for the Coulomb potential. As a result, either it is impossible to
obtain the solution for the next iteration step or some parameters of the
system are to be changed, for example, the density of the positive charge. The
proposed new iterative scheme for solving Kohn-Sham equations is freed up of
necessity to modify parameters of the system while iteration process.

\section{SELF-CONSISTENCY PROCEDURE}

The self-consistent electron density is constructed as the solution of the
Kohn-Sham equations.
\begin{equation}
\frac{1}{2}\nabla^{2}\psi_{\mathrm{E}}(z)+(E-u_{\mathrm{eff}}(z))\psi
_{\mathrm{E}}(z)=0\label{Eq(1)}%
\end{equation}%
\[
u_{\mathrm{eff}}=u(z)+u_{\mathrm{xc}}(z)
\]%
\begin{equation}
\nabla^{2}u=(N_{\mathrm{D}}-n(z))\label{Eq(2)}%
\end{equation}
Here $N_{\mathrm{D}}$ is the density of the positive background, $u$ is the
Coulomb potential energy of electron and $u_{\mathrm{xc}}$ is the
exchange-correlation potential energy, which we assume in the local-density approximation.

We represent the total electron density as a sum
\begin{equation}
n(z)=n_{\mathrm{ind}}(z)+n_{\mathrm{qu}}(z), \label{Eq(3)}%
\end{equation}
where induced electron density depends on the potential as follows
\[
n_{\mathrm{ind}}(z)=(\mu-u_{\mathrm{eff}}(z))^{3/2}.
\]
This expression is the usual quasi-classical solution of Eq.(\ref{Eq(1)})
averaged over Fermi wavelength scale and
\[
n_{\mathrm{qu}}(z)=2\sum_{\mathrm{E\leqslant E_{F}}}|\psi_{\mathrm{E}}%
|^{2}(z)-n_{\mathrm{ind}}(z)
\]
will be named as the quantum electron density.

Using Eq.(\ref{Eq(3)}) Poisson equation can be rewritten in a new form
\begin{equation}
\nabla^{2}u+n_{\mathrm{ind}}[u]=N_{\mathrm{D}}-n_{\mathrm{qu}}(z)\label{Eq(4)}%
\end{equation}
The total iteration scheme can be described as following
\begin{equation}%
\begin{array}
[c]{l}%
i=0,1,...,\text{ \ \ \ }n_{\mathrm{qu}}^{0}=0\\
\nabla^{\mathrm{2}}u^{i}+n_{\mathrm{ind}}(u^{i})=N_{\mathrm{D}}-n_{\mathrm{qu}%
}^{i}(z)\\
\frac{1}{2}\nabla^{\mathrm{2}}\psi_{\mathrm{E}}^{i}(z)+(E-u_{\mathrm{eff}}%
^{i}(z))\psi_{\mathrm{E}}^{i}(z)=0\\
n_{\mathrm{qu}}^{i+1}=\sum_{\mathrm{E\leqslant E_{F}}}|\psi_{\mathrm{E}}%
^{i}|^{2}(z)-n_{\mathrm{ind}}^{i}\left[  u_{\mathrm{eff}}^{i}\right]
\end{array}
\label{Eq(5)}%
\end{equation}
This procedure was tested in solving the self-consistent set of the equations
for two tasks within the jellium model:\newline 1. Semi-infinite electron gas
which is bounded by self-consistent potential\newline 2. Semi-infinite
electron gas which is bounded by infinite potential barrier.

\section{SEMI-INFINITE ELECTRON GAS BOUNDED BY SELF-CONSISTENT POTENTIAL}%

\begin{figure}
[h]
\begin{center}
\includegraphics[
trim=0.000000in 0.000000in 0.001308in -0.003618in,
height=11.9826cm,
width=17.2852cm
]%
{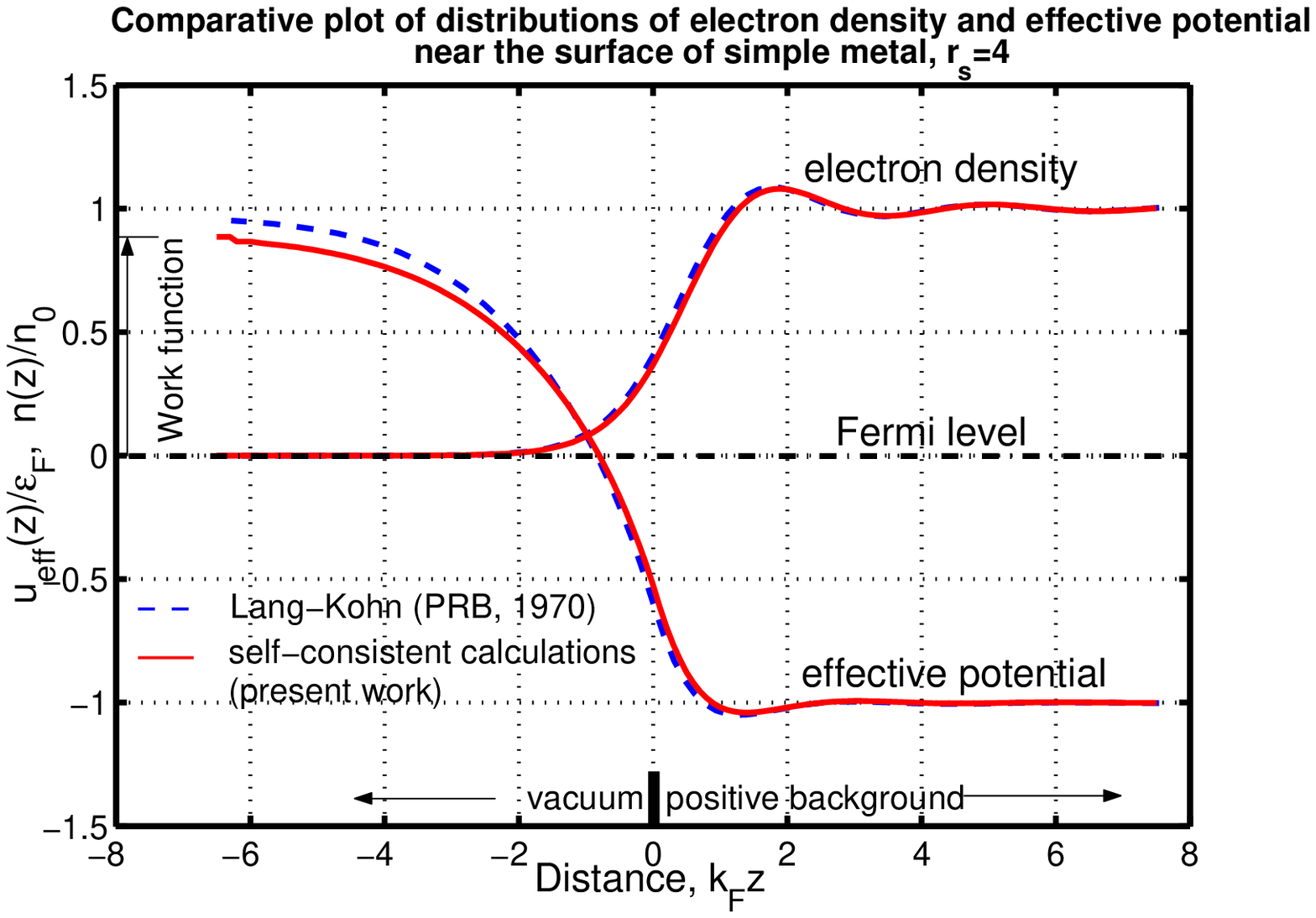}%
\caption{ }%
\label{Fig.1}%
\end{center}
\end{figure}
Using the local-density approximation for exchange and correlation potentials
we calculated the self-consistent distribution of the electrons and the work
function in the wide range of the bulk electron densities. All computations
were done in the framework of the jellium model, where the positive ions are
replaced by the uniform background of positive-charge density. This background
fills the half-space $z\geq0$ and is neutralized by the electron density.
Figure \ref{Fig.1} shows the distributions of electron density and effective
potential calculated self-consistently for $R_{\mathrm{s}}=4$. These curves
are compared with the corresponding results from \cite{LK-SE}, where the
straight iteration scheme was claimed not converging (see \cite{LK-SE},
Appendix B) and authors did have to use, in fact, variational method for
solving Eqs. (\ref{Eq(1)}-\ref{Eq(2)}). Our iteration scheme (\ref{Eq(5)})
allows to obtain the self-consistent answer through the straight solution of
Kohn-Sham equation set.

According to the well-known Budd-Vannimenus theorem \cite{BV73}, the
self-consistent potential must satisfy to the follow sum rule%
\begin{equation}
\Delta_{{ \mathrm{B}}}=u(0)-u(\infty)=\bar{n}\frac{d\epsilon
(\bar{n})}{d\bar{n}}. \label{BV-rule}%
\end{equation}
Here $\epsilon$ is an energy per one electron for the homogeneous electron
gas. Substituting the explicit expression for $\epsilon$, it is easy to
calculate $\Delta_{{ \mathrm{B}}}$ as the function of the electron
density. The same parameter obtained from our self-consistent solution we
denote as $\Delta_{{ \mathrm{SC}}}$. As it can be seen in Table I,
parameters $\Delta_{{ \mathrm{B}}}$ and $\Delta
_{{ \mathrm{SC}}}$ are in a good agreement. This fact is the
independent evidence that our solution is fully self-consistent \cite{BV73}.%
\begin{figure}
[h]
\begin{center}
\includegraphics[
natheight=6.816500in,
natwidth=6.371100in,
height=10.1638cm,
width=9.5048cm
]%
{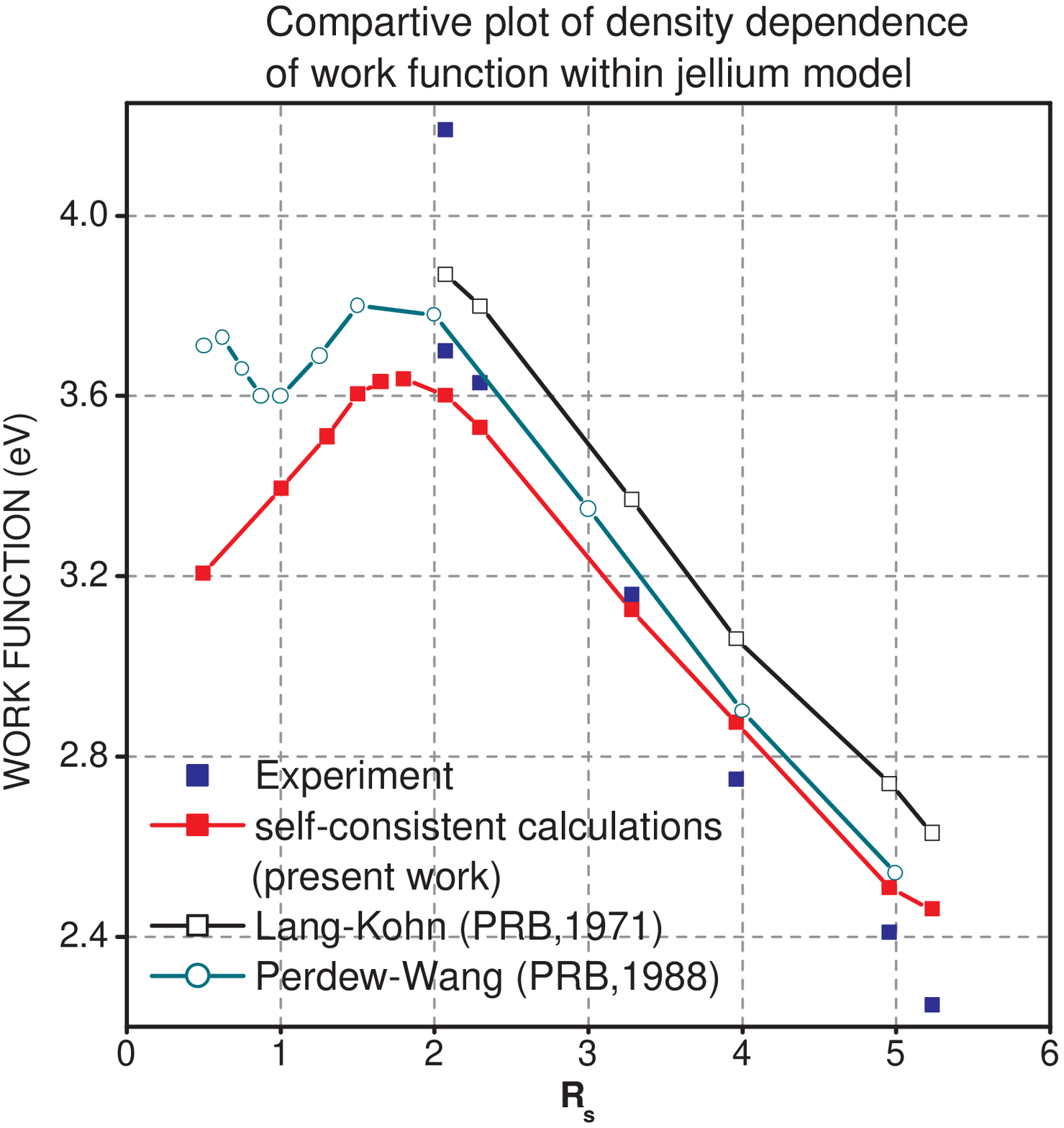}%
\caption{ }%
\label{Fig.2}%
\end{center}
\end{figure}
The work function is defined as
\[
W=u_{\mathrm{eff}}(-\infty)-\mu.
\]
The dependence of the self-consistently calculated
$W_{{ \mathrm{SC}}}$ on Wigner-Seitz parameter $R_{s}$ is shown in
Figure \ref{Fig.2} in companion with results of not fully self-consistent
calculations in \cite{LK-WF}-\cite{PW88}. This figure provides comparative
analysis of work function calculations obtained by different technique.\newpage

TABLE I. Electron density dependence of the work function\newline
\begin{tabular}
[c]{|l|c|c|c|}\hline
R$_{s}$ & W$_{{ \mathrm{SC}}}$(eV) & $\Delta
_{{ \mathrm{B}}}$ & $\Delta_{{ \mathrm{SC}}}$\\\hline
0.5 & 3.21 & 0.358 & 0.358\\
1.0 & 3.40 & 0.317 & 0.317\\
1.3 & 3.51 & 0.291 & 0.291\\
1.5 & 3.60 & 0.274 & 0.274\\
1.65 & 3.63 & 0.261 & 0.261\\
1.8 & 3.64 & 0.248 & 0.248\\
2.07 & 3.60 & 0.224 & 0.224\\
2.3 & 3.53 & 0.204 & 0.204\\
3.28 & 3.12 & 0.115 & 0.115\\
3.99 & 2.87 & 0.051 & 0.063\\
4.96 & 2.51 & -0.048 & -0.041\\
5.23 & 2.46 & -0.075 & -0.074\\\hline
\end{tabular}

\section{SEMI-INFINITE ELECTRON GAS BOUNDED BY INFINITE POTENTIAL BARRIER}

In this section we apply our iteration scheme for calculating self-consistent
electron density in the case of electron gas bounded by infinite potential
barrier. The positive charge and the electrons fill the half-space $z\geq0$.
The task was considered with accounting for the external electric field
perpendicular to the surface plane. The presence of the electric field
modifies the boundary condition for Coulomb potential at the surface. Using
the results of this task, the capacitance of a tunnel structure can be
estimated.%
\begin{figure}
[h]
\begin{center}
\includegraphics[
height=12.4922cm,
width=15.8706cm
]%
{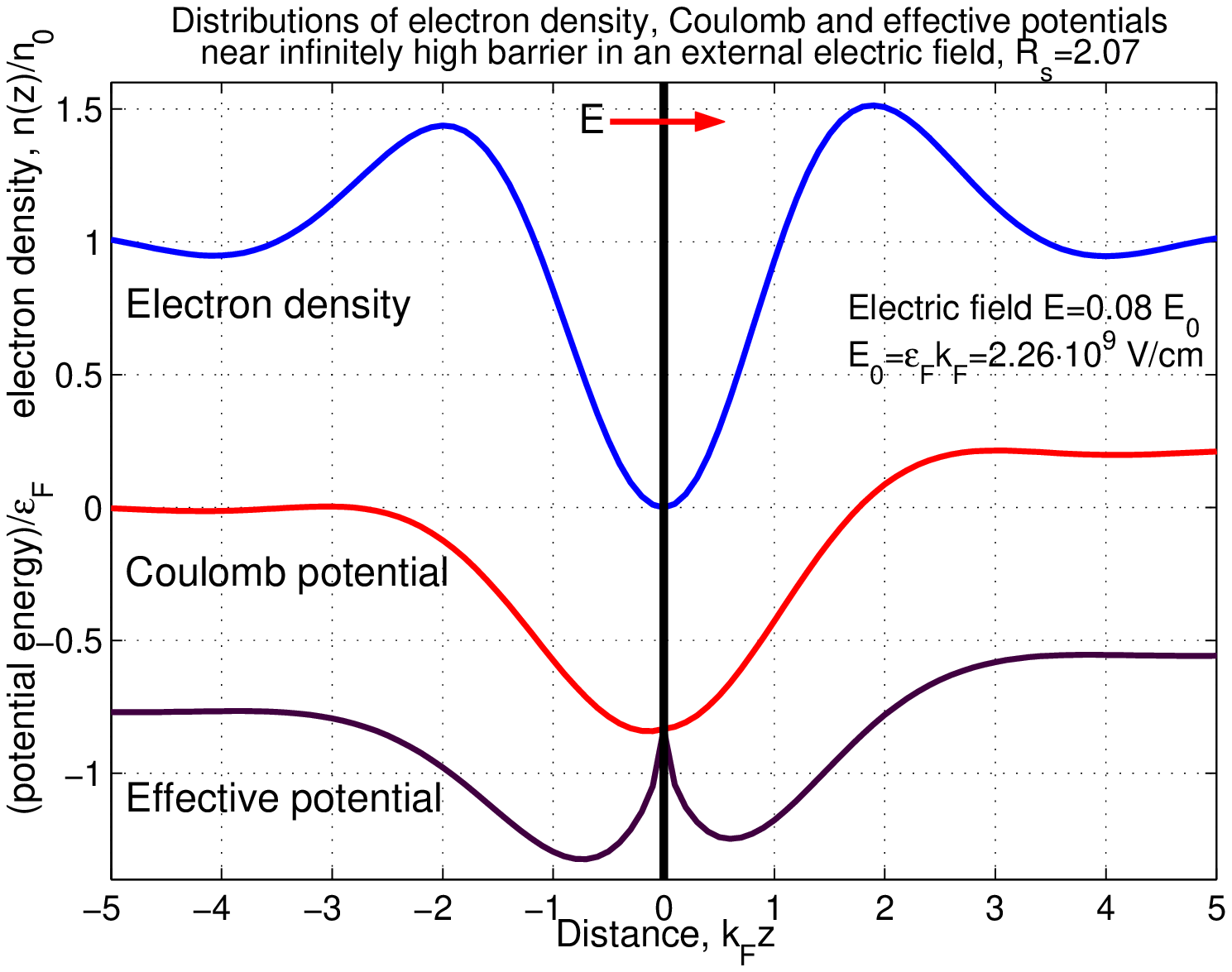}%
\caption{ }%
\label{Fig.3}%
\end{center}
\end{figure}
Figure \ref{Fig.3} demonstrates the distributions of electron density,
effective and Coulomb potentials for $R_{s}=2.07$ calculated by our technique.
The solution of this task can be used for an estimation of the capacitance of
a real finite-height barrier structure. The structure consists of two
$n-$doped semiconductor half-spaces, separated by a potential barrier of width
$d$. The total capacitance is $C=dQ/dV$, where $Q\,$is a total charge in the
semiconductor lead and $V=u(\infty)-u(-\infty)$ is the voltage drop over the
structure. Our calculations corresponds to the non-transparent infinitely high
barrier with $d=0$ and give us the maximum achievable specific capacitance in
barrier structures.%
\begin{figure}
[hh]
\begin{center}
\includegraphics[
natheight=6.776700in,
natwidth=6.454100in,
height=10.1616cm,
width=9.6783cm
]%
{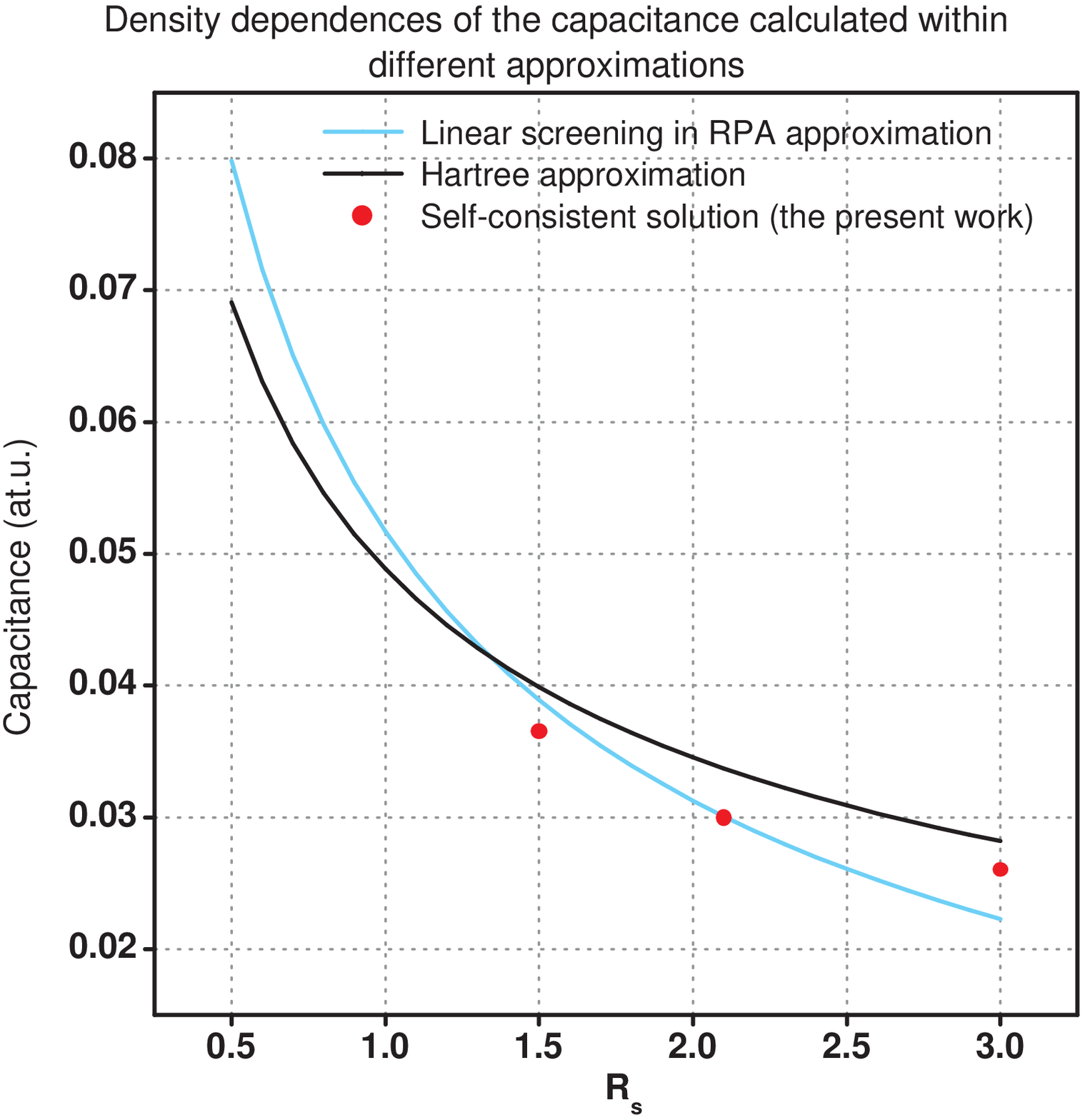}%
\caption{ }%
\label{Fig.4}%
\end{center}
\end{figure}
In Figure \ref{Fig.4} we plot the dependences of the capacitance on $R_{s}%
$calculated within the different approximations.

The partial support from the Russian Foundation for Basic Researches is
gratefully acknowledged.

\end{document}